\documentclass[letterpaper]{article} 
\usepackage[preprint]{aaai2027}  
\usepackage[hyphens]{url}  
\usepackage{graphicx} 
\usepackage{natbib}  
\usepackage{caption} 
\usepackage{algorithm}
\usepackage{algorithmic}
\usepackage{amsmath}
 \usepackage{multirow}
\usepackage{newfloat}
\usepackage{listings}
\DeclareCaptionStyle{ruled}{labelfont=normalfont,labelsep=colon,strut=off} 
\floatstyle{ruled}
\newfloat{listing}{tb}{lst}{}
\floatname{listing}{Listing}

\usepackage{booktabs}

\title{Direct or Mediated? Task-Dependent Audio Information Routing in Large Audio Language Models}

\author{
    Yizhou Zhang\textsuperscript{\rm 1},
    Wangjin Zhou\textsuperscript{\rm 1}\corresponding,
    Xin Gu\textsuperscript{\rm 2},
    Yichi Wang\textsuperscript{\rm 1},
    Wei Tan\textsuperscript{\rm 2},
    Yi Zhao\textsuperscript{\rm 2},
    Zhi Gong\textsuperscript{\rm 2},
    Keisuke Imoto\textsuperscript{\rm 1},
    Tatsuya Kawahara\textsuperscript{\rm 1}
}

\affiliations{
    \textsuperscript{\rm 1}Graduate School of Informatics, Kyoto University, Japan\\
    \textsuperscript{\rm 2}WXG, Tencent, China\\
    yizhang@sap.ist.i.kyoto-u.ac.jp
}

\begin{document}

\maketitle

\begin{abstract}
Large Audio Language Models (LALMs) have demonstrated strong performance across a wide range of audio understanding tasks. However, they are typically evaluated on single, coherent audio segments, leaving their behavior under less familiar input configurations underexplored. We study this issue through a controlled setting in which two audio segments are concatenated into a single input. Across multiple LALMs, we observe a striking task-dependent robustness gap: automatic speech recognition (ASR) remains comparatively stable, whereas audio question answering (AQA) degrades substantially. To investigate the mechanisms underlying this disparity, we analyze how audio information is routed through LALM decoders using layer-wise attention knockout. The results reveal distinct task-dependent pathways. ASR relies primarily on direct retrieval from audio tokens by answer tokens, whereas AQA depends more strongly on a mediated route in which audio information is first integrated into prompt tokens and subsequently accessed during generation. We further probe prompt-token representations under audio concatenation and find that task-relevant audio attributes remain readily decodable, particularly in middle and later decoder layers, even when AQA performance deteriorates sharply. This dissociation indicates that the failure cannot be explained by complete loss of audio information from the decoder states and is instead consistent with a downstream bottleneck in retrieving or utilizing prompt-mediated information during answer generation. Together, our findings reveal task-dependent audio information routing in LALMs and highlight information utilization as a potential limitation on their generalization.
\end{abstract}


\section{Introduction}

Large Audio Language Models (LALMs) extend the language modeling and reasoning capabilities of large language models (LLMs) to speech, environmental sounds, and music. Recent systems have demonstrated strong performance across a broad range of audio understanding tasks, including speech recognition and audio question answering \cite{xu2025qwen3,zhang2025mimo,tian2025step,ghosh2026audio}. Despite this progress, it remains unclear how audio information is represented, routed, and ultimately used inside LALM decoders.

This question is important because strong benchmark performance does not necessarily imply robust use of audio evidence. Most LALMs are evaluated on single, coherent audio segments drawn from familiar benchmark distributions, while their behavior under less conventional input configurations remains underexplored. Prior work has also shown that LALMs can exhibit insufficient audio grounding and hallucinated responses, particularly when the input departs from familiar settings \cite{cheng2026aha,zhao2026halluaudio,seth2026audio}. These limitations raise two broader mechanistic questions: through which internal information pathways does audio information influence the decoder’s predictions, and are these pathways shared across audio tasks or task-dependent?

We study this question by comparing two major tasks: automatic speech recognition (ASR) and audio question answering (AQA). ASR primarily requires the recovery of linguistic content, whereas AQA requires the model to identify evidence, preserve  attributes, and retrieve that evidence in response to a question. Whether these differences are reflected in distinct internal routing mechanisms remains unknown.

To expose this distinction, we introduce a controlled two-segment audio-concatenation setting. We concatenate two audio segments at the waveform level and evaluate the same models on ASR and AQA. Across four LALMs, we observe a consistent task-dependent robustness gap: ASR performance remains comparatively stable, whereas AQA performance deteriorates substantially.

We then use layer-wise attention knockout to trace the pathways through which audio information reaches generated outputs. The results reveal qualitatively different routing patterns. In ASR, direct access from answer tokens to audio-token representations is consistently important. In AQA, direct audio-to-answer sensitivity is comparatively weak, while audio-to-prompt and prompt-to-answer pathways have larger effects. These findings suggest that ASR and AQA differ in the relative importance of direct and mediated routes for accessing information from audio tokens.

Finally, we ask whether low two-segments AQA accuracy can be explained by the disappearance of task-relevant attributes from prompt-token representations. Layer-wise probing shows that these attributes remain linearly decodable, particularly in the middle and later decoder layers, even when end-task AQA accuracy is low. The failure therefore cannot be explained solely by a complete disappearance of the probed attributes from the decoder states.

Our contributions are threefold:

\begin{itemize}
    \item We introduce a controlled two-segment audio-concatenation setting and uncover a consistent task-dependent robustness gap across four LALMs: AQA performance is substantially lower in the concatenated multi-segment setting, while ASR remains comparatively robust.

   \item Through layer-wise attention knockout, we provide evidence for distinct relative pathway importance in the two tasks. ASR consistently exhibits strong direct audio-to-answer dependence, whereas AQA is more sensitive to prompt-mediated pathways than to direct audio-to-answer access.

    \item Through layer-wise probing, we show that audio attributes often remain linearly decodable from prompt-token representations. Together with the knockout results, this dissociation cannot be explained solely by a complete disappearance of the probed attributes from the decoder states.
\end{itemize}

\section{Related Work}

\subsection{Audio Understanding Tasks in LALMs}

Large Audio Language Models (LALMs) provide a unified generative framework for a diverse range of audio understanding tasks. In automatic speech recognition (ASR), recent systems typically project speech representations into the embedding space of a large language model and generate transcriptions autoregressively. Prior work has investigated cross-modal alignment, contextual conditioning, multilingual recognition, streaming inference, and robustness under real-world acoustic conditions
\cite{ma2024embarrassingly,bai2024seed,chen2024bestow,xu2025fireredasr,song2025index,shi2026qwen3}.

Audio question answering (AQA) places different demands on the model. Beyond recognizing acoustic or linguistic content, the model must identify task-relevant evidence and use it to produce an answer conditioned on a question. Research in this area has progressed from task-specific datasets and architectures
\cite{fayek2020temporal}
to broader benchmarks that evaluate open-ended audio understanding and reasoning
\cite{sakshi2025mmau,kumar2026mmau}.

Although both ASR and AQA require access to audio information, they have largely been studied as separate capabilities. Consequently, it remains unclear whether the two tasks rely on the same internal mechanisms for transferring audio information to generated outputs. Our work addresses this question by comparing their information-routing pathways within the same LALM decoders.

\subsection{Robustness of LALMs}

Recent work has examined the robustness of LALMs from several complementary perspectives. AHa-Bench
\cite{cheng2026aha}
and HalluAudio
\cite{zhao2026halluaudio}
evaluate audio hallucinations across multiple domains, while AHA
\cite{seth2026audio}
and HALAS
\cite{baranski2026halas}
study adversarially induced and naturally occurring hallucinations, respectively. Other studies have identified limitations in temporal localization and long-audio understanding
\cite{wang2026listening},
as well as in processing multiple audio inputs and reasoning across clips
\cite{chen2024beyond,kumar2026polyaudio}.

These studies demonstrate that strong performance on conventional benchmarks does not necessarily translate into reliable behavior under more challenging input conditions. Most prior work characterizes such failures primarily at the behavioral level through benchmark construction and evaluation. In contrast, we use a controlled waveform-level concatenation setting to expose a task-dependent robustness gap between ASR and AQA, and then investigate the internal routing mechanisms associated with this difference.

\subsection{Mechanistic Analysis of Multimodal Models}

A growing body of work uses probing and causal interventions to study how multimodal models process and transmit information. Studies of vision-language models have revealed stage-wise computation across decoder layers and token positions, including perceptual grounding, cross-modal integration, task reasoning, and answer decoding
\cite{zhang2025crossmodal,yu2025how,nikankin2025different, fu2025hidden}.
In speech and audio-language models, probing, activation patching, and attention-based analyses have been used to characterize acoustic-to-semantic transformations and mechanisms associated with hallucinations, repetitions, modality conflicts, and factual retrieval
\cite{glazer2026beyond,glazer2026audio,cho2026wins}.

Building on this line of work, we examine whether ASR and AQA route and use audio information differently within the same model. We combine layer-wise attention knockout with prompt-representation probing to distinguish how information is transferred and utilized along different routes.

\section{Methods}
\label{sec:methods}

We analyze audio information flow in LALM decoders using three complementary
tools. First, controlled audio concatenation exposes task-dependent differences
in robustness. Second, layer-wise attention knockout estimates the functional
importance of selected information-transfer pathways. Third, layer-wise probing
tests whether task-relevant audio attributes remain decodable from prompt-token
representations. We partition the decoder sequence into audio tokens
\(\mathcal{A}\), prompt tokens \(\mathcal{P}\), and autoregressively generated
answer tokens \(\mathcal{Y}\).

\subsection{Audio Concatenation}

We construct a composite input by concatenating two audio segments
\(x^{(1)}\) and \(x^{(2)}\) at the waveform level:
\[
    x^{\mathrm{cat}} = x^{(1)} \Vert x^{(2)},
\]
where \(\Vert\) denotes waveform-level concatenation. We refer to
\(x^{(1)}\) and \(x^{(2)}\) as the front and back segments, respectively.

For ASR, the target transcription is the ordered concatenation of the
transcriptions associated with the two segments:
\[
    y^{\mathrm{cat}}
    = y^{(1)} \Vert y^{(2)}.
\]

For AQA, we construct three types of questions using attributes associated
with the two segments: (1) \textbf{content} questions ask the model to identify
the attributes corresponding to the two segments from a set of candidate
attributes that includes distractors; (2) \textbf{order} questions ask which
attributes occur in the front and back segments; and (3) \textbf{location}
questions ask in which segment or segments a target attribute occurs.

This design is intended to align the AQA evaluation with ASR under the same concatenated-audio setting while preserving the intrinsic differences between the two tasks. In both cases, the model receives two waveform-level concatenated audio segments and must recover information about their content, temporal order, and segment-wise position. Correctly transcribing concatenated audio requires the model to recognize the content of both segments, preserve the order in which they occur, and reflect their positions within the concatenated input. Accordingly, the three types of AQA questions explicitly evaluate content, order, and location. By keeping the input construction, constituent audio segments, and underlying information requirements aligned as far as possible, the comparison controls for major factors other than the task formulation itself. It therefore provides a more fine-grained assessment of whether the model can distinguish, align, and use information across the two constituent segments under ASR and AQA objectives.

\subsection{Attention Knockout}

We trace information flow by suppressing selected directed attention edges.
Let \(\mathcal{T}\) denote a set of target-token positions and
\(\mathcal{S}\) denote a set of source-token positions. For the attention
logit from source position \(j\) to target position \(i\) in layer \(\ell\)
and head \(h\), denoted by \(M_{ij}^{(\ell,h)}\), we define the intervened
logits as
\[
\widetilde{M}_{ij}^{(\ell,h)}
=
\begin{cases}
-\infty, & i \in \mathcal{T},\; j \in \mathcal{S},\\
M_{ij}^{(\ell,h)}, & \mathrm{otherwise}.
\end{cases}
\]

Setting the selected logits to \(-\infty\) removes the corresponding attention
edges after the softmax operation. The intervention is applied to a contiguous
sliding window of decoder layers, while all other attention connections remain
unchanged.

We examine three directed pathways:
\[
\mathcal{P} \leftarrow \mathcal{A},
\qquad
\mathcal{Y} \leftarrow \mathcal{A},
\qquad
\mathcal{Y} \leftarrow \mathcal{P}.
\]

These interventions estimate, respectively, the functional importance of
audio-to-prompt transfer, direct audio-to-answer retrieval, and
prompt-mediated retrieval during answer generation.

The reliability of attention knockout depends on the model achieving high
clean-inference accuracy on the evaluated test examples. If the model already
produces an incorrect prediction without intervention, the effect of
suppressing a particular pathway is difficult to interpret. We therefore
conduct attention knockout primarily on examples that are correctly and
reliably solved under clean inference.

For each pathway and layer window, the intervention effect is defined as the
degradation in task performance relative to clean inference. A larger
degradation provides stronger evidence that the suppressed pathway is
functionally important for the model's prediction.

\subsection{Layer-Wise Probing}

We use layer-wise probing to test whether task-relevant audio information is
decodable from prompt-token representations. At each decoder layer \(\ell\),
we mean-pool the hidden states of the prompt-token positions:
\[
    r_{\mathcal{P}}^{(\ell)}
    =
    \frac{1}{|\mathcal{P}|}
    \sum_{i\in\mathcal{P}} h_i^{(\ell)}.
\]
An independently trained probe \(g^{(\ell)}\) then predicts the corresponding
task-relevant audio attribute:
\[
    \hat{c}^{(\ell)}
    =
    g^{(\ell)}
    \left(r_{\mathcal{P}}^{(\ell)}\right).
\]

The probing accuracy measures the linear decodability of the target attribute from the prompt representation at each layer. Low probing accuracy provides evidence that the attribute is weakly represented or not linearly accessible. Conversely, high probing accuracy together with poor end-task performance shows that the attribute remains decodable.

\section{Experimental Setup}
\begin{table*}[t]
\centering
\small
\setlength{\tabcolsep}{3.6pt}
\label{tab:concat_results}
\begin{tabular}{lcccccccccc}
\toprule
& \multicolumn{2}{c}{SpeechCommands}
& \multicolumn{2}{c}{GTZAN}
& \multicolumn{2}{c}{ESC-50}
& \multicolumn{2}{c}{WER}
& \multicolumn{2}{c}{CER} \\
\cmidrule(lr){2-3}
\cmidrule(lr){4-5}
\cmidrule(lr){6-7}
\cmidrule(lr){8-9}
\cmidrule(lr){10-11}
Model
& Single & Concat.
& Single & Concat.
& Single & Concat.
& Single & Concat.
& Single & Concat. \\
\midrule
Audio-Flamingo-Next
& 96.67 & 43.67
& 93.83 & 35.33
& 97.67 & 54.33
& 1.68 & 1.86
& 0.49 & 0.57 \\
MiMo-Audio
& 94.67 & 40.67
& 85.17 & 45.00
& 95.17 & 53.33
& 3.19 & 4.49
& 1.49 & 2.80 \\
Step-Audio-R1
& 97.67 & 56.00
& 87.83 & 40.00
& 97.50 & 67.33
& 2.21 & 3.13
& 1.09 & 2.12 \\
Qwen3-Omni
& 98.00 & 88.67
& 96.00 & 82.67
& 97.67 & 83.33
& 1.43 & 1.53
& 0.39 & 0.54 \\
\bottomrule
\end{tabular}
\caption{
Performance of four LALMs on single-segment attribute-recognition controls and concatenated-audio AQA, together with single- and concatenated-input ASR. AQA results are multiple-choice accuracies (\%; higher is better), whereas LibriSpeech results are WER and CER (\%; lower is better).
}
\label{tab:concat_results}
\end{table*}

\subsection{Models and Implementation}
\label{sec:models_and_implementation}

We evaluate four Large Audio Language Models (LALMs): Audio-Flamingo-Next-Instruct, MiMo-Audio-Instruct, Step-Audio-R1, and Qwen3-Omni-Instruct. We first evaluate all four models under the controlled audio-concatenation setting to compare their robustness on automatic speech recognition (ASR) and audio question answering (AQA). We then conduct attention-knockout and layer-wise probing analyses on all four models to examine how audio information is routed and represented within their decoders.

For each model, we use its official audio preprocessing pipeline. All audio
clips are resampled to 16~kHz before concatenation and model-specific
preprocessing. All experiments are conducted on NVIDIA H20 GPUs.

\subsection{Concatenated-Audio Evaluation Setting}
\label{sec:concat_setup}

We follow the waveform-level audio-concatenation procedure defined in
the \textbf{Methods} section. The two constituent clips are referred to as the
\textbf{front} and \textbf{back} segments. Both clips are first resampled to
16~kHz and then concatenated directly at the waveform level. No additional
silence or explicit boundary marker is inserted between the two segments.

\paragraph{Concatenated ASR.}

Utterance pairs from the LibriSpeech test-clean split are concatenated at
the waveform level. The reference transcription is formed by placing the
transcript of the front utterance before that of the back utterance. We report
word error rate (WER) and character error rate (CER).

For the single-segment control, the constituent utterances are presented to
the model individually using the same decoding configuration. The reference
for each input is its original utterance-level transcription.

\paragraph{Concatenated AQA.}

We construct multiple-choice AQA examples from SpeechCommands
\cite{warden2018speech}, GTZAN \cite{tzanetakis2002musical}, and ESC-50
\cite{piczak2015esc}, representing spoken commands, music genres, and
environmental sounds, respectively. Each example consists of two source clips
and a question of one of the following three types:

\begin{itemize}

    \item \textbf{Content.}
    Given a set of candidate attributes containing distractors, the model
    identifies the two attributes that match the content of the front and back
    segments. The attributes are command words for SpeechCommands, music genres
    for GTZAN, and sound categories for ESC-50. Each answer is an attribute set
    containing the two attributes that occur in the two segments.

    \item \textbf{Order.}
    Given two candidate attributes, the model identifies which
    attributes occur in the front and back segments. The attributes are command
    words for SpeechCommands, music genres for GTZAN, and sound categories for
    ESC-50.

    \item \textbf{Location.}
    Given a target attribute, the model determines whether it occurs in both
    segments, only the front segment, only the back segment, or neither
    segment. The target attributes are command words for SpeechCommands, music
    genres for GTZAN, and sound categories for ESC-50.

\end{itemize}

The answer classes are balanced within each question type. We report
multiple-choice accuracy averaged over the three question types.

\paragraph{Single-Segment Control.}

To distinguish failures of basic attribute recognition from failures caused
by multi-segment processing, we also present every constituent clip
individually. For each clip, the model answers a four-option question about
its command word, music genre, or sound category for SpeechCommands, GTZAN,
or ESC-50, respectively. The position of the correct option is randomized.

\begin{figure*}[t!]
    \centering
    \includegraphics[
        width=\textwidth,
        height=0.78\textheight,
        keepaspectratio
    ]{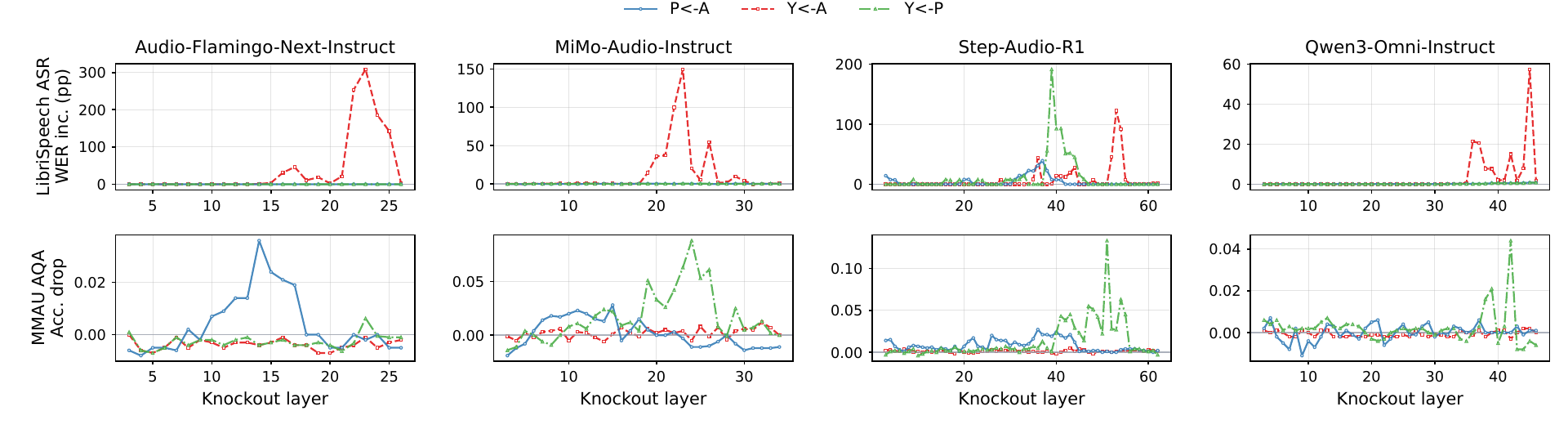}
    \vspace{-2mm}
    \caption{
    Layer-wise effects of attention knockout on three directed pathways
    across decoder depth. The top row reports the increase in WER on
    LibriSpeech ASR, and the bottom row reports the decrease in
    multiple-choice accuracy on MMAU test-mini AQA. Each point corresponds
    to a contiguous five-layer intervention window indexed by its starting
    layer. Positive values indicate performance degradation relative to
    clean inference.
    }
    \label{fig:knockout}
    \vspace{-3mm}
\end{figure*}

\subsection{Attention-Knockout Setting}
\label{sec:knockout_setup}

We apply the attention-knockout intervention defined in the
\textbf{Methods} section using a contiguous five-layer sliding window over the
decoder depth. Within each window, the selected directed attention edges are
suppressed, while all other attention connections remain unchanged. Each
pathway is intervened on independently.

We evaluate the interventions on two complementary tasks: automatic speech
recognition (ASR) and audio question answering (AQA). For ASR, we use
LibriSpeech and report word error rate (WER) and character error rate (CER).
We select LibriSpeech because it is a standardized speech-recognition benchmark
with high-quality transcriptions and a relatively controlled speech-only
setting. This controlled setting reduces variation unrelated to the linguistic
information flow from audio to generated text, making the corresponding
attention pathways more stable and the effects of pathway suppression easier
to interpret. In addition, the model achieves reliable clean-inference
performance on LibriSpeech, providing a meaningful baseline against which
intervention-induced increases in WER and CER can be measured. Reporting both
metrics captures the intervention effect at complementary levels of
granularity: WER measures word-level recognition errors, whereas CER is
sensitive to finer-grained transcription errors.

For AQA, we use the MMAU test-mini split and report multiple-choice
accuracy. We adopt MMAU for the mechanistic analysis because it covers a
broader range of audio content than the individual datasets used in the
concatenated-audio evaluation. This diversity allows us to examine whether
the candidate information pathways remain functionally important across
different forms of acoustic evidence rather than within a single audio
domain.

Moreover, the evaluated models achieve relatively high and stable
clean-inference accuracy on MMAU test-mini. This is important for the
attention-knockout analysis introduced in the \textbf{Methods} section,
because the effect of suppressing a pathway is more interpretable when the
model can reliably solve the unmodified examples. Under such conditions,
performance degradation is more likely to reflect the functional importance
of the disrupted information pathway rather than baseline prediction errors
or unstable task behavior. The multiple-choice format also provides a
well-defined evaluation target, enabling consistent measurement of the
accuracy drop caused by each intervention.

We examine the three directed pathways defined in the \textbf{Methods}
section:
\[
    \mathcal{P} \leftarrow \mathcal{A},
    \qquad
    \mathcal{Y} \leftarrow \mathcal{A},
    \qquad
    \mathcal{Y} \leftarrow \mathcal{P}.
\]

For a metric \(m\) for which higher values indicate better performance, such
as AQA accuracy, we define the intervention effect as
\[
    \Delta m
    =
    m_{\mathrm{clean}}
    -
    m_{\mathrm{knockout}}.
    \label{eq:delta_higher}
\]
For error metrics for which lower values indicate better performance, such as
WER and CER, we reverse the subtraction:
\[
    \Delta m
    =
    m_{\mathrm{knockout}}
    -
    m_{\mathrm{clean}}.
    \label{eq:delta_lower}
\]
Under both definitions, a positive value indicates performance degradation
caused by the intervention.

We apply each intervention to every valid contiguous five-layer window. For
each pathway and window, we report the performance change relative to clean
inference, yielding a depth-resolved estimate of the pathway's functional
importance.

\subsection{Layer-Wise Probing Setting}
\label{sec:probe_setup}

Following the procedure defined in the \textbf{Methods} section, we freeze the
LALM and train an independent linear probe at each decoder layer using the
mean-pooled prompt-token representation. We construct probing examples from
the same three datasets used in the concatenated-AQA evaluation:
SpeechCommands, GTZAN, and ESC-50. The target attributes are command words,
music genres, and sound categories, respectively.

For concatenated inputs, we train separate linear probe heads to predict the
attributes of the front and back segments. We train separate probes for each
model, decoder layer, dataset, segment position, and input condition. The
corresponding single-segment inputs are used as a control.

We partition the source clips into training, validation, and test sets using
a 60\%/20\%/20\% split. This partitioning is performed before constructing
concatenated examples, and audio pairs are formed only from clips belonging
to the same split. We enforce speaker-level disjointness for SpeechCommands,
track-level disjointness for GTZAN, and original-recording-level
disjointness for ESC-50. Consequently, no speaker, source clip, or
constituent of a concatenated pair is shared across the training, validation,
and test sets. The single-segment and concatenated-input conditions use the
same underlying source-clip partitions.

Before training each probe, we standardize its input representations using
the feature-wise mean and variance estimated from the training split. Each
probe consists of a single affine classification layer and is trained using
the cross-entropy objective. We optimize the probes with AdamW using a
learning rate of \(1\times10^{-3}\), a batch size of 256, and a maximum of
50 epochs. We apply early stopping with a patience of five epochs and retain
the checkpoint achieving the highest validation accuracy. The LALM remains
frozen throughout training, and gradients are computed only for the probe
parameters.

We report top-1 classification accuracy at each decoder layer. For
concatenated inputs, we compute accuracy separately for the front- and
back-segment probes and report their macro-average:
\[
    \mathrm{Acc}_{\mathrm{macro}}
    =
    \frac{
        \mathrm{Acc}_{\mathrm{front}}
        +
        \mathrm{Acc}_{\mathrm{back}}
    }{2}.
\]


\section{Experiments}
\label{sec:experiments}

\begin{figure*}[t!]
  \centering
  \includegraphics[
      width=\textwidth,
      height=0.78\textheight,
      keepaspectratio
  ]{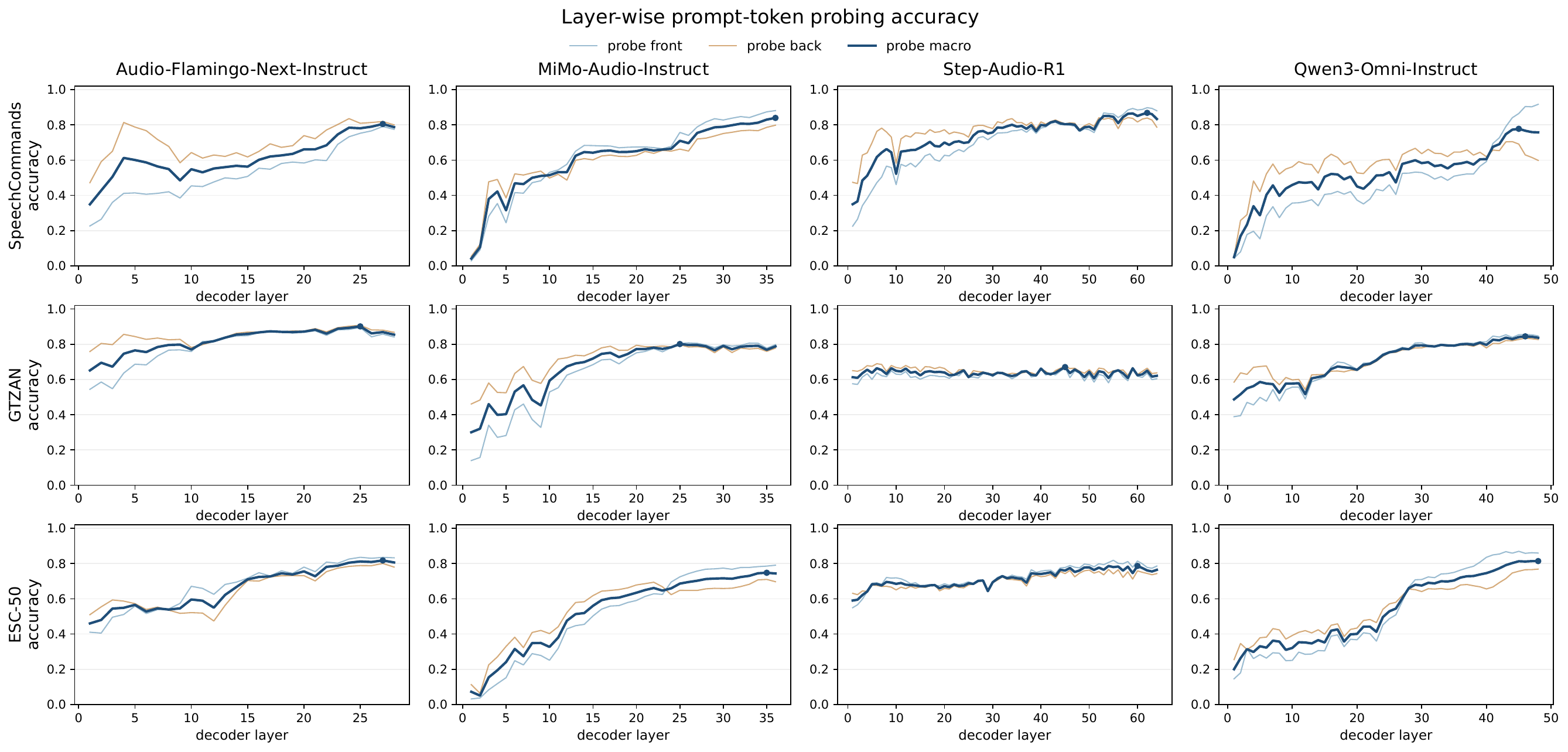}
  \caption{
  Layer-wise linear probing accuracy for task-relevant audio attributes
  decodable from prompt-token representations under concatenated inputs.
  Rows correspond to SpeechCommands, GTZAN, and ESC-50, and columns
  correspond to the evaluated LALMs. For each decoder layer, separate
  probes predict the attributes of the front and back segments; their
  macro-average is also reported. Higher accuracy indicates greater
  linear decodability of the target attribute.
  }
  \label{fig:probing}
\end{figure*}


\subsection{Concatenation Reveals a Task-Dependent Robustness Gap}
\label{sec:behavioral_results}

Table~\ref{tab:concat_results} reveals a consistent asymmetry between the two
task formulations. Under ASR, concatenating two utterances introduces only a
small absolute increase in transcription error: across the four models, WER
rises by 0.10--1.30 percentage points and remains below 4.5\%. The linguistic
content of both segments therefore remains largely recoverable after
waveform-level concatenation.

The AQA results show a different pattern. The single-segment controls confirm
that all four models can recognize the relevant command, genre, or sound
category in isolation, with accuracies between 85.17\% and 98.00\%. However,
performance is substantially lower when the model must answer questions about
two concatenated segments. For Audio-Flamingo-Next, MiMo-Audio, and
Step-Audio-R1, concatenated-input accuracy falls to 35.33--67.33\% across the
three datasets. Qwen3-Omni is more robust, retaining 82.67--88.67\% accuracy,
but still performs below its single-segment controls on every dataset.

These results support two conclusions. First, the lower AQA performance cannot be attributed simply to a lack of basic attribute-recognition ability, because all four models recognize the relevant attributes accurately when the constituent clips are presented individually. Second, although concatenated ASR and the three AQA question types place closely aligned demands on the model—recovering the content of both segments, preserving their order, and distinguishing their segment-wise locations—they exhibit markedly different robustness under the same waveform-level concatenation setting. The models can reliably transcribe the two segments in sequence, yet often fail when the same segment-level information must be explicitly identified and used to answer content, order, or location questions.

This discrepancy establishes the central behavioral phenomenon studied in the remainder of the paper. The relevant information requirements are shared across the two task formulations, but successful recovery of that information in ASR does not translate into equally reliable use of it in AQA. This raises a mechanistic question: why do tasks requiring closely related audio information exhibit such different outcomes? We next investigate whether this divergence reflects task-dependent differences in the internal pathways through which audio information reaches and influences the generated answer.

\subsection{ASR and AQA Exhibit Different Patterns of Pathway Sensitivity}
\label{sec:routing_results}

Figure~\ref{fig:knockout} shows that the behavioral asymmetry is accompanied by distinct patterns of pathway sensitivity. In ASR, preventing answer tokens from directly attending to audio tokens generally produces the most substantial performance degradation. Overall, the evaluated models exhibit a strong dependence on direct audio-to-answer access for ASR, although the peak layer and effect magnitude vary across architectures. In Step-Audio-R1, the prompt-to-answer pathway also shows a pronounced effect, but this does not alter the main observation that direct audio-to-answer access plays an important role in ASR.

AQA exhibits a different ordering of pathway importance. Compared with disrupting direct audio-to-answer access, intervening on the integration of audio information into prompt-token states or restricting the subsequent influence of prompt-token states on answer generation generally produces larger decreases in accuracy. Overall, AQA is more sensitive to pathways involving prompt-mediated computation and less sensitive to interventions on the direct audio-to-answer pathway. Although the effect magnitudes and depth profiles vary substantially across models, this relative tendency is broadly consistent across the evaluated architectures.

The contrast therefore lies not only in the overall magnitude of the intervention effects, but also in the relative importance of different computational pathways. ASR depends more strongly on direct access from answer tokens to audio tokens, whereas AQA relies more heavily on audio-conditioned prompt representations and their subsequent influence on answer generation. The integration of audio information into prompt-token states tends to show sensitivity in earlier layers, while prompt-to-answer interactions are often more sensitive in later layers. This pattern is consistent with a temporally ordered, prompt-mediated computation.

Attention knockout does not uniquely identify an information-carrying circuit, as removing an attention edge also renormalizes the remaining attention weights, and a small or negative intervention effect does not demonstrate that the corresponding pathway carries no information. Nevertheless, attention knockout remains an effective method for causally evaluating the relative importance of candidate information pathways. We therefore interpret Figure~\ref{fig:knockout} as evidence that the two tasks assign different relative importance to the evaluated pathways, rather than as proof of exclusive routing mechanisms. This interpretation motivates the next question: when AQA fails under concatenation, has the relevant audio information disappeared from the prompt-token states, or does it remain available but fail to influence the final answer?

\subsection{Prompt Representations Retain Linearly Decodable Audio Information}
\label{sec:probing_results}

Figure~\ref{fig:probing} examines whether information about the constituent
audio segments can be linearly decoded from prompt-token representations.
Across most model--dataset pairs, probing accuracy increases through the early
and middle decoder layers and remains relatively high in later layers. This
pattern indicates that information about the front and back audio segments is
present in a linearly decodable form within the prompt states across a
substantial portion of the decoder depth.

On SpeechCommands, the best macro-averaged probing accuracies are 80.5\%,
84.0\%, and 86.8\% for Audio-Flamingo-Next, MiMo-Audio, and Step-Audio-R1,
respectively. On GTZAN, the corresponding best probing accuracies are 90.1\%,
80.1\%, and 67.0\%, while on ESC-50 they are 81.8\%, 74.8\%, and 78.7\%.
Qwen3-Omni achieves best probing accuracies of 77.8\%, 79.6\%, and 81.4\% on
SpeechCommands, GTZAN, and ESC-50, respectively.

These results show that command, genre, and sound-category attributes can be
linearly recovered from prompt-token representations in multiple models and
across different decoder layers. The probing analysis therefore provides
evidence that prompt states preserve information about the constituent audio
segments in a form accessible to a linear classifier.

These probing results show that task-relevant audio attributes remain present in prompt-token representations even when the model fails to produce the correct answer, and are particularly linearly decodable in the middle and later decoder layers. Therefore, the decline in AQA performance cannot be explained simply by a complete disappearance of the relevant audio information from the decoder states.

\section{Conclusion}
\label{sec:conclusion}

We investigated task-dependent audio processing in four Large Audio Language Models. Under two-segment waveform concatenation, ASR remained comparatively robust, whereas multi-segment AQA performance was substantially lower than single-segment attribute-recognition controls.

Attention knockout revealed different patterns of pathway sensitivity: ASR showed stronger dependence on direct audio-to-answer access, while AQA was generally more sensitive to audio-to-prompt and prompt-to-answer pathways. Probing under concatenated inputs further showed that task-relevant attributes often remained linearly decodable from prompt-token states. Together, these findings reveal task-dependent differences in behavioral robustness, pathway sensitivity, and representational decodability.

\section{Limitations}
\label{sec:limitations}

First, the attention-knockout analysis is indirect with respect to the concatenation failures examined in this work. Because interpretable intervention effects require correct and stable clean predictions, we apply knockout on standard benchmark examples that the models solve reliably. The results therefore characterize pathways supporting successful prediction, but do not directly identify where computation fails in incorrect concatenated-AQA cases. In addition, attention knockout cannot strictly localize information routing to specific layers or components, since removing attention edges perturbs the surrounding computation and renormalizes the remaining attention weights. We thus interpret the results as evidence about the relative importance of candidate pathways, rather than as a precise diagnosis of the failure mechanism or a unique circuit localization.

Second, we focus on a controlled two-segment waveform-concatenation setting. While this design facilitates systematic analysis, it represents only a simplified form of multi-audio input and may not capture the challenges arising in more realistic scenarios. Future work should test whether the observed patterns generalize to overlapping audio, background noise, longer recordings, additional segments, and more complex cross-segment interactions.

\bibliography{aaai2027}

\end{document}